# An Adaptive Overcurrent Protection for Solar-based DC Microgrids Using IEC 61850

Saeed Sanati, Maher Azzouz, and Ahmed Awad, *Senior Members, IEEE*

*Abstract*— Over-Current (OC) protection is one of the pervasive protections in solar-based DC microgrids. Fast operation is a key advantage of its popularity. On the other hand, utilizing OC in DC microgrids has some challenges that are not in AC grids. Some of these challenges are related to the grounding approach of the DC microgrid, and others are related to the high rise time of the fault current in DC microgrids. Considering these challenges, an adaptive OC scheme with high selectivity and speed is presented in this paper. The proposed scheme is communication-assisted and relies on IEC 61850 protocol. In this scheme, different setting groups for each OC relay are defined, and based on the grid and fault conditions, a setting group is selected. This option is performed considering the data transferred via communication level using IEC 61850 protocol between relays. To evaluate the efficiency of the proposed scheme, simulations using MATLAB software and the experimental tests using OPAL-RT real-time simulator and Zenon software are presented.

## I. Introduction

MICROGRIDS are local networks that have energy consumers and producers. They can work in grid-connected or islanded mode. Thus, microgrids should be controlled and protected locally. Microgrids are divided into AC and DC categories. DC microgrids are becoming increasingly prevalent due to the rise in demand for DC power, which includes traction systems, office equipment, household appliances, lighting equipment, as well as DC power sources such as fuel cells, solar photovoltaic (PV) panels, and DC energy storage devices [1, 2]. DC microgrids do not encounter certain issues that are present in AC grids, such as synchronization, frequency control, reactive power control, and harmonics [3].

On the other hand, DC microgrids exhibit distinct fault characteristics from AC grids due to various factors such as the low impedance in the DC system, the short length of transmission lines and cables, the presence of the DC link capacitor, and the grounding arrangement. Fault currents in DC microgrids have a large amplitude and short rise time. Therefore, it reaches its maximum value rapidly as short as 2 $ms$. Additionally, grounding in DC microgrids may result in high fault impedance, leading to a reduction in fault current. Hence, fault detection by Over-Current (OC) protection is challenging [4]. Furthermore, current transformers' saturation in DC microgrids exacerbates the fault detection issue [5-7]. During high current DC faults, instrument transformers, especially current transformers, are prone to mislead the protection relays. High transient fault currents in DC microgrids can seriously damage converters, cables, transmission lines, circuit breakers (CBs), and other network equipment [8, 9].

One of the most common protection schemes for fault isolation in distribution networks is OC. However, due to the need for high operation speed in DC microgrids, protection coordination of OC relays is challenging. OC protection is divided into two schemes, directional and bidirectional. Bidirectional OC protects both the area in front and back of the relay, whereas directional OC only protects the area in front of the relay. Directional OC is more common considering better performance in terms of selectivity rather than bidirectional OC. The OC scheme assumed in this paper is directional. Fig. 1 illustrates the coordination of two downstream and upstream OC relays. If a fault occurs, as shown in Fig. 1(a), the downstream relay should trigger the trip command, and the upstream relay should not operate so that less outage occurs in the grid. In other words, the relays should be coordinated, and the protection system should work with high selectivity. In this case, assuming the definite-time scheme, the settings of the OC relays should be adjusted, as shown in Fig. 1(b). The time difference between $t_1$ and $t_2$ ensures that the relays operate selectively.

In DC microgrids, due to the need for high-speed fault clearance, maintaining a minimal time difference $\Delta t = t_2 - t_1$ is crucial, even if it exacerbates coordination due to the potential for high-impedance faults arising from the unique grounding configuration. Reducing the time differences $\Delta t$ may disturb the coordination of the protection system and causes the reduction of the protection selectivity, thus increasing unnecessary outages.

The timing of fault clearance due to the operation of the relay $R_2$ is shown in Fig. 2. Considering this timing, the minimum time required for coordination, $\Delta t_{min}$, should be obtained according to (1) for selective protection. In (1), $\Delta t_{min}$ is based on the periods demonstrated in Fig. 2. The tripping relay receives the command trip from the OC relay and amplifies the trip signal to energize the CB trip coil. $t_{TR}$ is defined as the trip relay delay as displayed in Fig. 2. $t_{CBop}$ is the required time for the operation of the mechanical parts of CB. $t_{Arc}$ is the delay such that any arc between the CB contacts is distinguished. $t_{reset}$ is the time required to reset the OC relay as depicted in Fig. 2.

According to [10-13], $t_{TR}$, $t_{CBop}$, $t_{Arc}$ and $t_{reset}$ are around

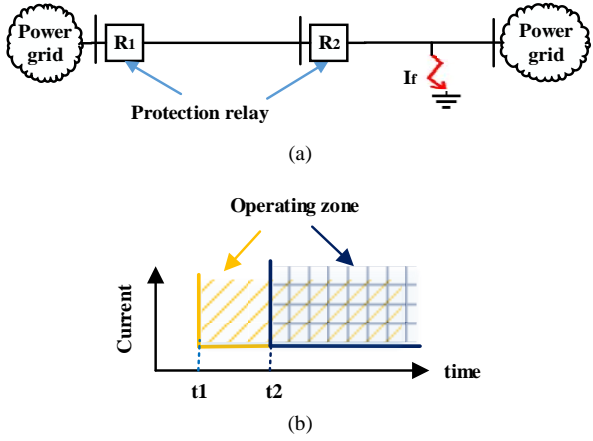

Fig. 1. OC relay placement and coordination, (a) upstream and downstream relay placement, and (b) Relay coordination considering definite-time characteristics.

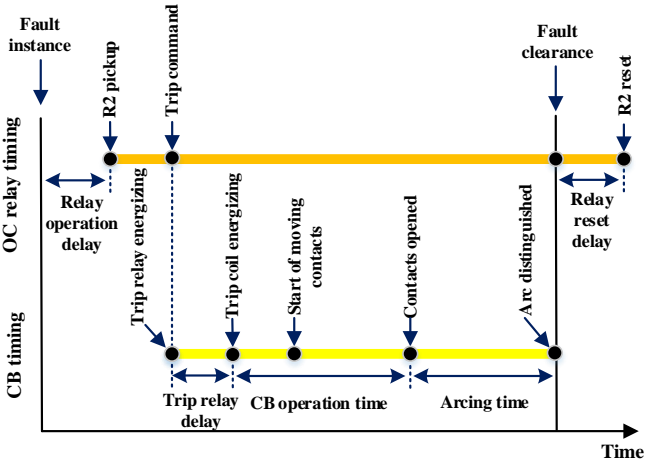

Fig. 2. OC relay and CB timing during the fault condition.

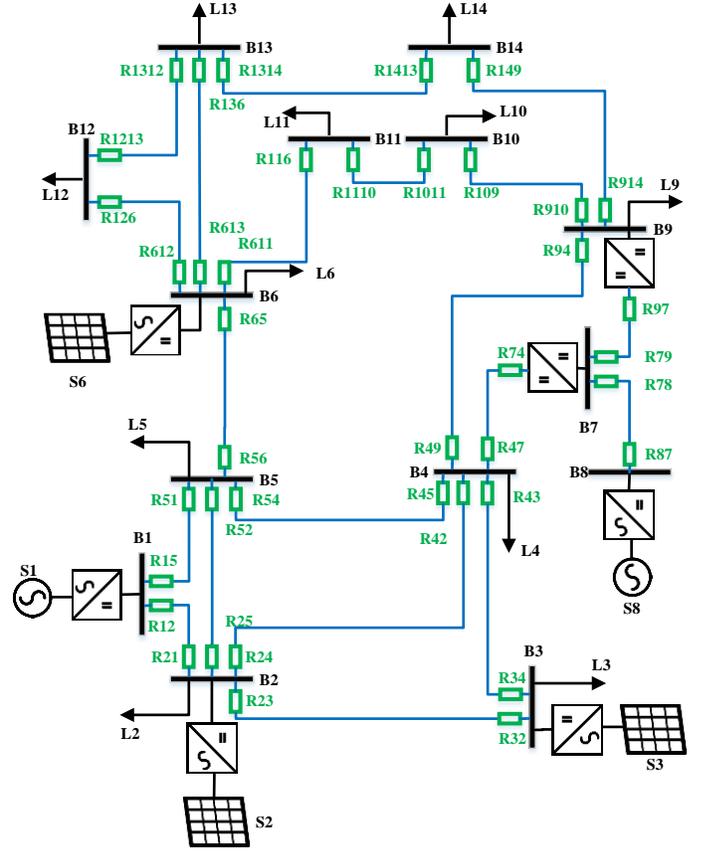

Fig. 3. Single line diagram of the investigated DC microgrid.

4 ms, 15 ms, 5 ms, and 5 ms, respectively, for DC circuit breakers and OC relays with the standard inverse scheme. Therefore, $\Delta t_{min}$ to ensure selectivity is equal to 29 ms.

$$\Delta t_{min} = t_{TR} + t_{CBop} + t_{Arc} + t_{reset} \quad (1)$$

In solar-based DC microgrids, the peak fault current can occur within a very short duration of just 2 ms. Consequently, the traditional coordination method in DC microgrids faces a challenge in balancing the need for rapid protection operation with coordination requirements.

The protection of DC microgrids using OC relays was discussed in [1]. This method is fast and communication-assisted, but it can only be applied to power grids with a radial topology. In [14] and [15], OC protection was implemented for a DC microgrid with rectifiers that can limit fault currents. However, implementing such a protection method in more complex DC microgrid architectures may result in longer fault clearance times or unnecessary disconnections of larger parts of the grid during faults [10]. In [16], a framework was proposed based on the integration of unit-based protection, which has high sensitivity, speed, and selectivity, but it has low sensitivity to high-impedance faults [10]. [17] proposed a method based on adding a parallel LC filter to each pole to have resonance at a specific frequency during faulty conditions. Then, a Discrete Wavelet Transform (DWT) was utilized to extract this frequency for fault identification. Despite the ability to operate in low fault current magnitudes, that method requires additional elements and is not fast enough to clear the high rise time faults. [18] proposes an adaptive OC scheme for DC microgrids which is statistical-based. That method is efficient under various operating scenarios, including instantaneous switching operations of sources or loads, where fault impedance varies during the fault, however, it is applicable only on the ring type grids.

In this paper, an adaptive OC scheme for solar-based DC microgrids is presented, which is suitable for all grid architectures and does not need adding extra power equipment in the DC microgrid. Also, there is no need to have rectifiers with the ability to limit the fault current in this method. In the proposed method, according to clustering DC microgrid status and fault conditions, several setting groups are defined for each relay. At any moment, only one of the setting groups, which guarantees the highest speed and the best selectivity, is activated for a relay. The selection of the setting group is done through the information exchanged between the protection relays at the communication level. Network conditions and the status of other relays are shared between relays through the communication level. The communication protocol at the communication level is IEC61850. Based on this protocol, the required information is exchanged in the form of Generic Object-Oriented Substation Event (GOOSE) messages.

The proposed method is implemented on an interconnected DC microgrid, which consists of consumers, PV-based DGs, and storage devices. The investigated grid is shown in Fig. 3. This grid is a developed network based on the 14-bus IEEE benchmark network [17, 19]. An experimental test was conducted to assess the effectiveness of the proposed method, and the results are presented alongside simulation results.

## II. PRINCIPLE OF THE PROTECTION COORDINATION IN THE PROPOSED METHOD

For protection coordination in the proposed method, several setting groups are defined for OC relays. The number of these setting groups depends on the DC microgrid topology, while the setting parameters for each group depending on the DC microgrid conditions and possible faults. Choosing which setting settings group to be used by a relay is done via the exchange of information at the communication level. In case of a fault, each OC relay that has been picked up checks the pickup information of other relays. Then, it decides to issue a trip command instantaneously or after a delay $\Delta t$ according to (1). For this purpose, two types of information are used. Information about the status of DC microgrid switches and information about the operation of adjacent OC relays. In the following subsections, the effect of the microgrid operation status, the operation of the adjacent OC relays, and protocol configuration are investigated.

### A. Effect of the grid operation condition

The condition of grid operation includes the grid topology, location and sizes of loads and generators, energy storage devices, state of CBs and disconnector switches, state of the system grounding in the DC microgrid, impedances and lengths of transmission lines and cables, and condition of connection point to the upstream network.

In the proposed method, different operating conditions are checked in terms of the fault current. For conditions in which the minimum fault current differs by less than 10%, the same setting group is considered. To explain the proposed method, the solar-based DC microgrid shown in Fig. 3 is considered, which represents a modified version of the standard 14-bus IEEE grid [17, 19]. The method implementation on the R12 relay is explained as an example. The locations of sources, consumers, energy storage devices, OC relays, and CBs are demonstrated in Fig. 3, and the parameters of the investigated DC microgrid are given in Table I.

The investigated DC microgrid voltage is $\pm 750$ VDC for all areas (except buses $B7$ and $B8$), considering EU LVDC6002/95/EC guidelines [20]. For $B7$ and $B8$, the voltage level is 380 VDC, which is one of the most common DC voltage levels [21]. The grid is connected to the AC systems at two points, i.e., $B1$ and $B8$. $B1$ is a slack bus with the capacity of 1.5 MVA, and a 300 kVA synchronous generator is connected to $B8$. The voltage source converters at $B1$ control the DC grid voltage level and operate at unity power factor. Other sources in the investigated grid are solar PV generations with the capacity of 500 kW, 400 kW, and 300 kW placed at $B6$, $B3$, and $B2$, respectively. Solar PV plants are connected to the grid via DC/DC boost converter. The solar-based power sources include PV arrays made of twenty series-connected modules per string. The number of parallel strings is related to the source capacity. The maximum power of each module is 200 W and its voltage at the maximum power point is 29 V. The grounding system for the investigated DC microgrid is TN-S, the same as the grounding scheme recommended by most guidelines [17].

Usually, the OC relay's pickup current is set between twice the nominal line load and half of the minimum fault current in their protection zones [22]. There is a delay between the pickup and trip command issued by the OC relay, as displayed in Fig. 2. This time is related to the OC protection scheme and its configuration and setting. The inverse definite minimum time (IDMT) is the most common scheme for OC relays. The pickup, drop, and trip timings in this scheme are according to (2) [23, 24].

$$t = T \times \left[ \frac{k}{\left(\frac{I}{I_s}\right)^\alpha - 1} + L \right] \quad (2)$$

where, $t$ is the total operation time, $k$ and $\alpha$ and $L$ are factors following Table II, $I$ is the measured current, and $I_s$ is the pickup value, which is commonly considered the minimum fault current value, and $T$ is the time multiplier setting varying from 0.025 to 1.5 [23, 24].

TABLE I
GRID SPECIFICATIONS

| | | | | | |
|---|---|---|---|---|---|
| Grid power sources | S1 apparent power | 1 MVA | DC lines | L12,L23,L34,L45,L24,L25, L15, L56 and L49 | 5 km |
| | S2 active power | 300 kW | | L47, L612, L613, L611, L1213, L1314, L149, L109, L1011,L97 and L87 | 2 km |
| | S3 active power | 400 kW | | | |
| | S6 active power | 500 kW | | | |
| | S8 apparent power | 200 kVA | | Cables resistance | 0.018 Ω/km |
| Grid load consumptions | L2, L3, L4 and L5 consumption | 150 kW | Overall | Cables inductance | 3.2 mH/km |
| | L9, L10 and L11 consumption | 130 kW | | Grounding system | TN-S |
| | L12, L13 and L14 consumption | 110 kW | | Voltage | +380, +750, -750 VDC |

TABLE II
THE IDMT CURVE PARAMETERS

| Type of IDMT curve | Standard | K factor | α factor | L factor |
|---|---|---|---|---|
| Short time inverse | AREVA | 0.05 | 0.04 | 0 |
| Standard inverse | IEC | 0.14 | 0.02 | 0 |
| Very inverse | IEC | 1.5 | 1 | 0 |
| Exteremely inverse | IEC | 80 | 2 | 0 |
| Long time inverse | AREVA | 120 | 1 | 0 |
| Short time inverse | C02 | 0.023 | 0.02 | 0.016 |
| Moderately inverse | ANSI/IEEE | 0.051 | 0.02 | 0.011 |
| Long time inverse | C02 | 9.95 | 2 | 0.18 |
| Very inverse | ANSI/IEEE | 19.61 | 2 | 0.49 |
| Exteremely inverse | ANSI/IEEE | 28.2 | 2 | 0.12 |
| Rectifier protection | RECT | 45900 | 5.6 | 0 |

The OC scheme is non-unit protection; thus, it does not have a defined operation zone. However, for the best selectivity, the relays should work in such a way that the fault range is isolated and the fault is cleared with the minimum amount of outage. Therefore, the primary zone of each relay usually has a small overlap with the primary zone of the upstream or downstream relays. For example, the selective primary zones for relays $R12$, $R23$, $R24$, and $R25$ are shown in Fig. 4, while the secondary zone for each OC relay is the same as the primary zone of its downstream relay. For example, the $R12$ secondary zone is the same as the primary zones of $R23$, $R24$, and $R25$.

According to the protection zone of $R12$, the minimum fault current will occur at the end of the primary protection zone and the longest distance from the $R12$ relay. To consider all the possible conditions for the state of the switches in the DC microgrid in Fig. 3, the minimum fault current values are obtained by modeling the grid in Simulink/MATLAB and simulating the conditions, as presented in Table III. The inverters in simulations are made by connecting a thyristor D bridge and a thyristor Y bridge. The inverters are equipped with an internal DC fault and low-voltage AC voltage protection. Double contingencies for line outages and source outages are also considered in Table III.

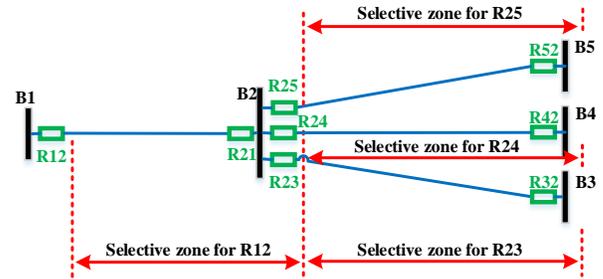

**Fig. 4.** Selective primary zones for R12 and its downstream OC relays considering investigated DC microgrid.

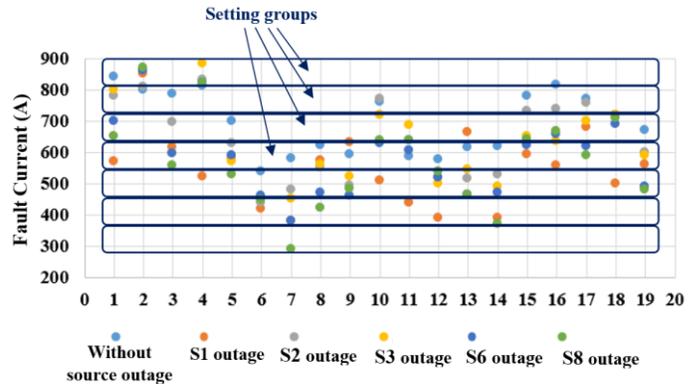

**Fig. 5.** Different setting groups defined considering Table III.

TABLE III
THE MINIMUM FAULT CURRENT VALUES

| No | Line outages | Sources outages | | | | | |
|---|---|---|---|---|---|---|---|
| | | Without outage | S1 outage | S2 outage | S3 outage | S6 outage | S8 outage |
| 0 | L12 | N/D | N/D | N/D | N/D | N/D | N/D |
| 1 | L15 | 840.3 | 570.3 | 780.3 | 800.3 | 700.3 | 650.3 |
| 2 | L25 | 799.1 | 849.1 | 809.1 | 859.1 | 859.1 | 869.1 |
| 3 | L24 | 785.5 | 615.5 | 695.5 | 595.5 | 595.5 | 555.5 |
| 4 | L23 | 811.5 | 521.5 | 831.5 | 881.5 | 821.5 | 821.5 |
| 5 | L34 | 699.5 | 579.5 | 629.5 | 569.5 | 589.5 | 529.5 |
| 6 | L45 | 538.9 | 418.9 | 438.9 | 448.9 | 458.9 | 448.9 |
| 7 | L56 | 579.6 | 379.6 | 479.6 | 449.6 | 379.6 | 289.6 |
| 8 | L47 | 621.5 | 571.5 | 551.5 | 561.5 | 471.5 | 421.5 |
| 9 | L49 | 591.5 | 631.5 | 491.5 | 521.5 | 461.5 | 481.5 |
| 10 | L78 | 758.5 | 508.5 | 768.5 | 718.5 | 628.5 | 638.5 |
| 11 | L79 | 586.4 | 436.4 | 636.4 | 686.4 | 606.4 | 636.4 |
| 12 | L612 | 577.8 | 387.8 | 517.8 | 497.8 | 517.8 | 537.8 |
| 13 | L613 | 614.4 | 664.4 | 514.4 | 544.4 | 464.4 | 464.4 |
| 14 | L611 | 618.9 | 388.9 | 528.9 | 488.9 | 468.9 | 368.9 |
| 15 | L1213 | 780.9 | 590.9 | 730.9 | 650.9 | 620.9 | 640.9 |
| 16 | L1314 | 815.9 | 555.9 | 735.9 | 635.9 | 655.9 | 665.9 |
| 17 | L914 | 768.3 | 678.3 | 758.3 | 698.3 | 618.3 | 588.3 |
| 18 | L910 | 718.4 | 498.4 | 718.4 | 718.4 | 688.4 | 708.4 |
| 19 | L1011 | 668.9 | 558.9 | 598.9 | 588.9 | 488.9 | 478.9 |

In Fig. 5 and according to Table III, all the values of the minimum fault currents are categorized into different groups, called setting groups. The minimum fault current is the fault current peak value. In the traditional scheme for OC relays, only a minimum value of the fault current is one of the main parameters of the relay settings. That value, in the traditional scheme, is the minimum value among all the values in Table III. However, in the proposed adaptive method, the required number of setting groups is defined instead of a single value. Each setting group represents the minimum fault currents that differ by less than 10% and is obtained by experiment. Therefore, according to Table III and Fig. 5, seven setting groups are defined for $R12$. The highest minimum fault current is 881 $A$. The difference between two minimum fault currents in consecutive groups should not exceed 10%; thus, the width of each setting group is defined as 85 $A$. This setting group width leads to defining the seven setting groups shown in Fig. 5. Considering the grid operation condition, one of the setting groups in the relay is active at any time. For example, the first setting group is active in one of the following conditions:
1- Line L15 is out and all sources are operating, or
2- Line L25 is out and S8 is disconnected, or
3- Line L25 is out and S1 is disconnected, or
4- Line L23 is out and S8 is disconnected, or
5- Line L23 is out and S3 is disconnected

*B. Effect of adjacent relay operation :*

In the proposed method for a fast and selective operation, instead of coordinating relays with a definite time, each OC relay uses the operation information of other OC relays to make decisions. For relays that are in the same direction as each other, the downstream relays share their operation information with the upstream relays. For relays in opposite directions, each relay shares its operation information with all relays and uses the information of downstream relays. It also uses the operation information of the adjacent relay in the opposite direction and its upstream relays. For example, considering Fig. 3, relay $R12$ uses the operation information of $R23$, $R24$, and $R25$ as its downstream relays. The operation information is shared contentiously before and during the fault occurrence and includes the pickup and drop status of those relays. Also, $R12$ will use the operation information of relay $R21$ as the opposite direction adjacent relay and the operation information of relays $R32$, $R42$, and $R52$ as upstream relays of relay $R21$. Each relay will issue an instantaneous trip if it is picked up and the opposite direction adjacent relay is picked up, according to Fig. 6. The proposed method is applied to radial and meshed microgrids. Fig. 6 is a simplified radial topology to display the faults that may be detected by $R1$.

In general, in the proposed method, each relay that is picked up works according to the following instructions:
1) If the opposite direction adjacent relay is picked up, it issues an instantaneous trip command. Such a situation is described in Fig. 6(a).
2) If downstream same-direction and downstream opposite-direction relays are picked up, a trip command will be issued after time $\Delta t$ according to (1). Such conditions are described in Figs. 6(b) and 6(c).
3) Other possible conditions will be considered as protection failure of other relays and the instantaneous trip command will be issued. As an example, a protection failure is detected if none of the downstream opposite-direction relays are picked up.

The flowchart of the proposed method is shown in Fig. 7. At the beginning, the OC relay sends its status into the communication network by GOOSE messages and receives the status of other OC relays and the operational condition of the DC microgrid. Then, the setting group is selected regarding the received information. Afterward, the relay employs the measured current to determine if it is greater than the setting group threshold or not. If this criterion is passed and the opposite direction adjacent relay is picked up, the process moves to the next step. In the next step, an instantaneous trip is issued if the downstream same-direction relay is picked up. However, the trip command is sent after a delay only if the opposite-direction relay is picked up.

*C. Relay communications and protocol configuration*

The IEC 61850 protocol provides a base for high-speed peer-to-peer communication. This protocol improves the control and protection communication in substations, power lines, power plants, and microgrids without imposing requirements for installing extra equipment [25, 26]. IEC 61850 can be implemented on a vast majority of protection systems that rely on the communication infrastructure and are able to interoperate and communicate with other elements [27].

GOOSE and Sample Value (SV) messages are two main specific communication service mappings defined in IEC 61850. The SV messages are employed for the acquisition of raw values measured by instrument transformers, sensors, or measuring units. This service transfers digitized instantaneous quantities into multicast Ethernet frames, such as primary currents and voltages [25].

GOOSE messages are unidirectional, time-critical, and multicast. Multicast messaging property allows all Intelligent Electronic Devices (IEDs) to send a GOOSE message that can carry both binary and analog values, although they are primarily used to indicate changes in the state of parameters, like CB states. Considering VLAN priority, these messages are used for fast protection and control information exchange between IEDs. GOOSE service is suitable for time-critical applications

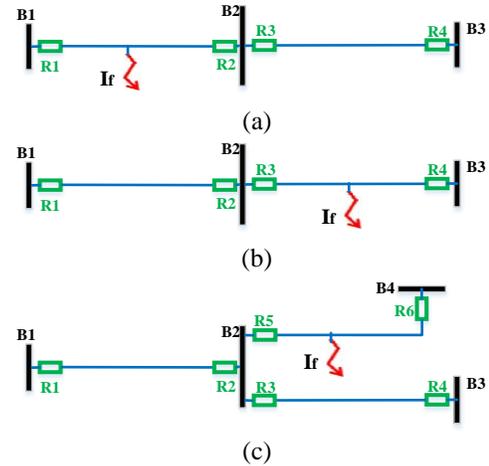
Fig. 6. Different possible fault conditions sensed by $R1$ relay.

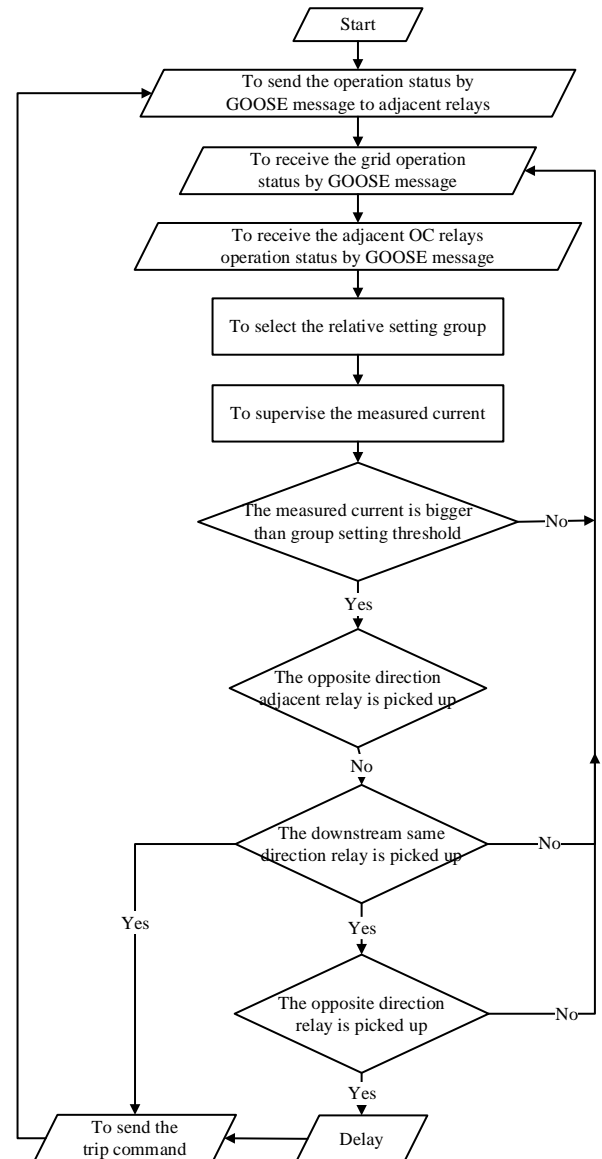
Fig. 7. Proposed Algorithm.

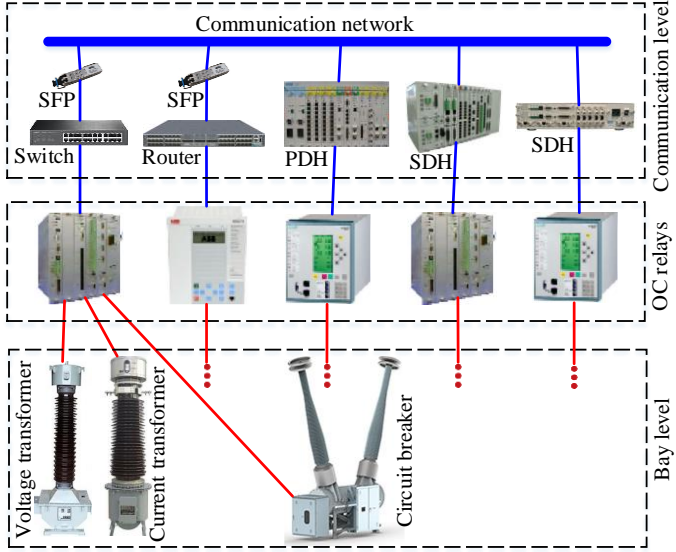

**Fig. 8.** Communication configuration.

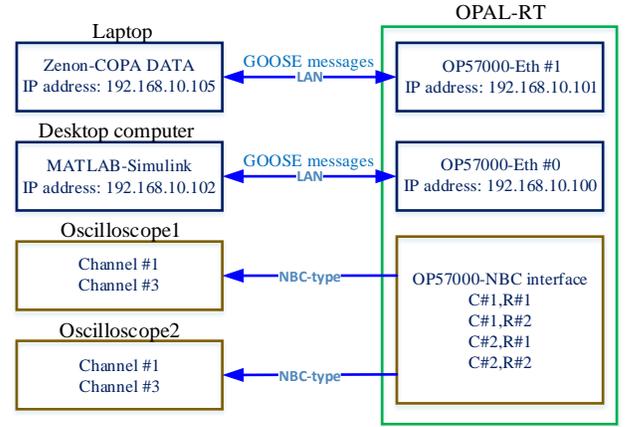

**Fig. 9.** The configuration of the test bed.

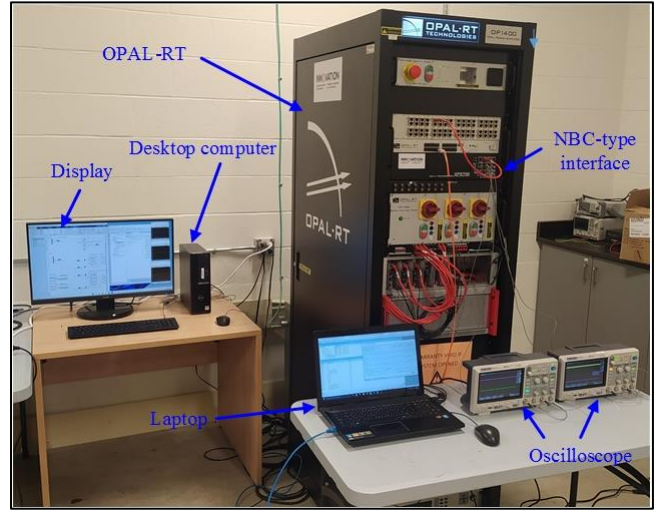

**Fig. 10.** The photograph of the implemented setup.

such as protection functions [25]. In the proposed method for transferring GOOSE messages, the communication configuration is displayed in Fig. 8.

High-speed operation requirement leads to utilizing high-speed communication infrastructure in the proposed method. Considering the short line length in DC microgrids, fiber optic is a common choice for communication networks. The minimum proposed speed is 100 $Mbps$, which is quite available with fiber optics active equipment, like network switches, routers, Synchronous Digital Hierarchy (SDH), and Plesiochronous Digital Hierarchy (PDH). The GOOSE message transferring delay, considering a 100 $Mbps$ communication network, is as short as $0.066\ ms$. However, this delay could be up to $1.8\ ms$ considering the chosen security algorithm [28]. Application of the security algorithm is not mandatory according to the IEC61850 standard. The communication level in the proposed method is isolated from external communication networks. Therefore, to increase the communication speed, security algorithms are not implemented on the GOOSE packets.

## III. REAL-TIME SIMULATIONS OF THE PROPOSED METHOD

The efficiency of the proposed method is evaluated with real-time simulations, using OPAL-RT OP1400 real-time simulator and MATLAB. To simulate transferring the GOOSE messages, i.e., related to the adjacent OC relays and grid operation states, Zenon automation software is used. The configuration of the test bed for experimental results is displayed in Fig. 9, and the photograph of the implemented setup is shown in Fig. 10.

The test grid for real-time simulations is shown in Fig. 3, with details given in Table I. The proposed algorithm in Fig. 7 is implemented for relay $R12$ using Zenon on a laptop. It should be noted that the proposed algorithm should be applied for all of the OC relays in the DC microgrid and relay $R12$ is a sample for case study in our experimental tests. The DC microgrid operation status and other OC relays' operation statuses are modeled using MATLAB on a desktop computer and simulated using a real-time simulator.

The studied scenarios are made by changing the condition of the DC microgrid. The information of these changes, including grid operation status and OC relays' operation statuses, are transferred by GOOSE messages from OPAL-RT to $R12$, which is simulated on a laptop. Afterward, the information of $R12$ operation, if exists, is sent to OPAL-RT by GOOSE messages, and the DC grid status is updated according to that information.

In the real-time simulations, according to Fig. 5, seven setting groups for $R12$ are defined. Deferent studied scenarios include pole-ground and pole-pole short circuit occurrence on $L12$ and every adjacent power line while every single power generation source in the DC microgrid is absent or present. Regarding fault occurrence on $L12$'s adjacent power lines, two cases are tested, (a) correct protection operation and (b) protection failure. In the first case, $R12$'s adjacent OC relays operate before the operation of $R12$, while in the second case, $R12$ operates in its secondary zone for backup protection because of the failure of its adjacent OC relays. Considering these two case studies, Tables IV and V show the OC operation times for the proposed method in comparison with the standard inverse scheme for pole-pole faults.

TABLE IV
R12'S OPERATION TIME FOR THE PROPOSED METHOD AND STANDARD INVERSE SCHEME IN PARENTHESES IN THE CASE OF CORRECT ADJACENT RELAY OPERATION

| Fault location | Sources outages | | | | | |
|---|---|---|---|---|---|---|
| | Without outage | S1 outage | S2 outage | S3 outage | S6 outage | S8 outage |
| L12 | 0.51 ms (4.52 ms) | 0.68 ms (4.21 ms) | 0.58 ms (3.25 ms) | 0.62 ms (2.85 ms) | 0.56 ms (2.94 ms) | 0.64 ms (3.98 ms) |
| L23 | N/D | N/D | N/D | N/D | N/D | N/D |
| L24 | N/D | N/D | N/D | N/D | N/D | N/D |
| L25 | N/D | N/D | N/D | N/D | N/D | N/D |

TABLE V
R12'S OPERATION TIME FOR THE PROPOSED METHOD AND STANDARD INVERSE SCHEME IN PARENTHESES IN THE CASE OF ADJACENT RELAY FAILURE

| Fault location | Sources outages | | | | | |
|---|---|---|---|---|---|---|
| | Without outage | S1 outage | S2 outage | S3 outage | S6 outage | S8 outage |
| L12 | N/D | N/D | N/D | N/D | N/D | N/D |
| L23 | 0.69 ms (21.11 ms) | 0.75 ms (32.19 ms) | 0.83 ms (26.22 ms) | 0.81 ms (30.12 ms) | 0.77 ms (29.94 ms) | 0.66 ms (33.88 ms) |
| L24 | 0.89 ms (29.79 ms) | 0.91 ms (21.18 ms) | 0.95 ms (31.92 ms) | 1.01 ms (32.01 ms) | 0.85 ms (24.52 ms) | 0.79 ms (34.35 ms) |
| L25 | 0.92 ms (30.88 ms) | 0.87 ms (27.27 ms) | 0.94 ms (31.89 ms) | 0.67 ms (25.94 ms) | 0.84 ms (29.78 ms) | 0.96 ms (33.11 ms) |

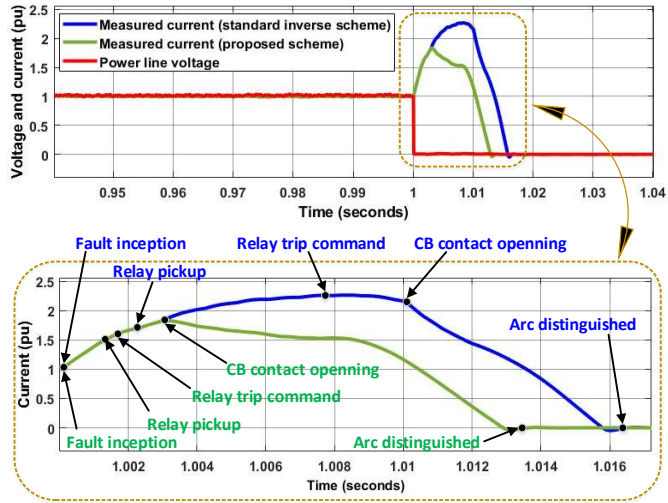

**Fig. 11.** The current waveform and the OC operation sequences in the case of correct adjacent relay operation while pole-pole fault occurs at $L12$ without any power source outage.

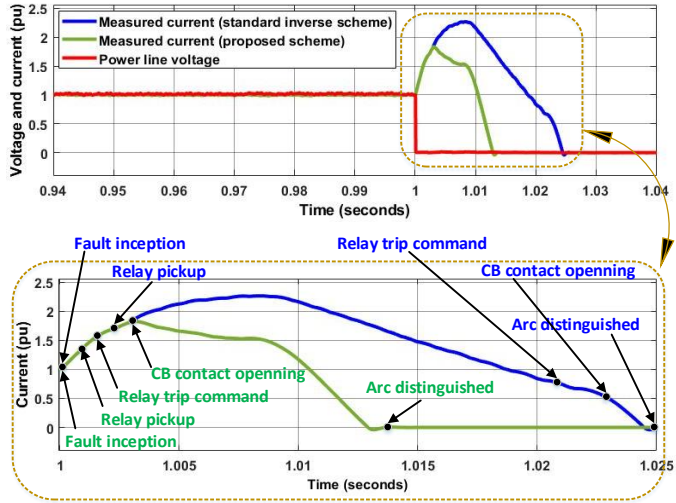

**Fig. 12.** The current waveform and the OC operation sequences in the case of adjacent relay failure while pole-pole fault occurs at $L23$ without any power source outage.

For fault scenarios in these tables, the relay operation times of the proposed method are much lower than those of the standard inverse scheme. As shown in Table IV, $R12$ correctly waits for other relays to operate in the fault scenarios where the fault location is in the secondary zone of the relay. However, if the protection failure is detected considering the proposed method, $R12$ operates faster than the standard inverse scheme to clear the fault, as displayed in Table V.

The current waveform seen by $R12$ and the operation sequences according to the proposed method for the first case study, i.e. pole-pole fault occurs on $L12$ without any power source outage, is shown in Fig. 11. Also, the standard inverse scheme sequences are displayed comparatively in Fig.11. The effect of the inverters, converters and their protection functions is considered by modeling the converters/inverters internal topology and their internal protections including DC fault and low voltage AC voltage protections. The time of the CB and OC relay actions are displayed in Fig. 11. The relay pickup point is displayed in Fig. 11 for the proposed method versus the standard inverse scheme which occurs after fault inception. It can be seen that the relay pickup s faster for the proposed method. Samely, the relay trip command is issued faster in the proposed method rather than in the standard inverse scheme. The whole process of relay operation for the proposed method is 3.01 ms faster than the standard inverse scheme, regarding the timing displayed in Fig. 11. As depicted in Fig. 11 and Table IV, the proposed method is faster in fault detection/clearing as compared to the standard inverse scheme.

For the case of adjacent protection failure, i.e., pole-pole fault occurring on $L23$ without any power source outage, the current waveform and the OC operation sequences are depicted in Fig. 12. As it is shown in Fig. 12, the fault clearance using the proposed method is 13 $ms$ faster than the standard inverse scheme. It can be seen considering the timing sequence related to the relay operation and CB action, displayed in Fig. 12.

The real-time simulations are repeated considering a pole-ground fault, and the results are illustrated in Figs. 13-14. The condition of the DC microgrid is kept similar to that of the cases depicted in Figs. 11-12, where a fault occurs at $L12$ while all power lines and power sources are operational. Fig. 13 displays the current waveform and the OC relay sequences in the case of correct adjacent relay operation. Fig. 14 shows those waveforms in the case of adjacent relay failure with the same prerequisites. According to Figs. 13-14, the fault clearance with the proposed is 3 $ms$ and 12 $ms$ faster than that with the standard inverse scheme for the protection failure and protection normal operation, respectively.

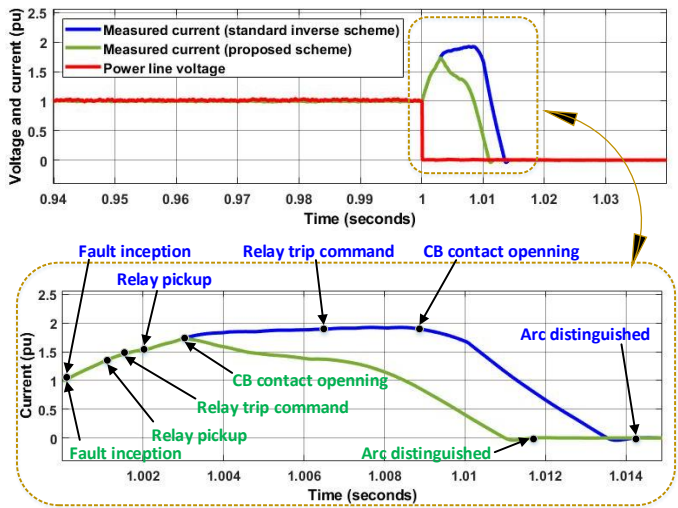

**Fig. 13.** The current waveform and the OC operation sequences in the case of correct adjacent relay operation while pole-ground fault occurs at *L*12 without any power source outage.

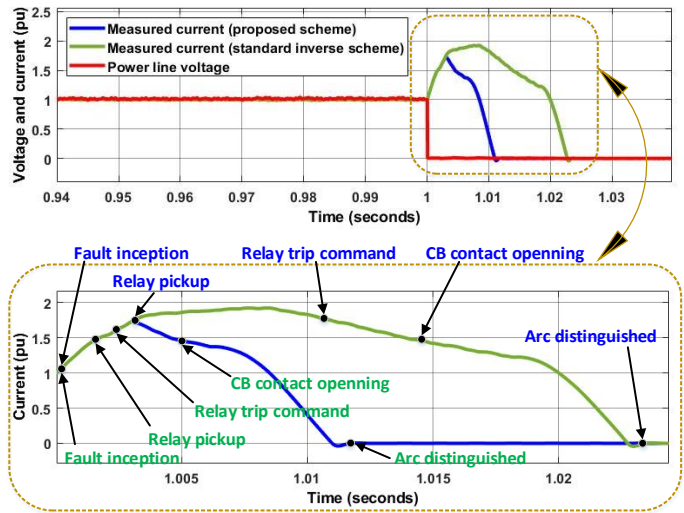

**Fig. 14.** The current waveform and the OC operation sequences in the case of adjacent relay failure while pole-ground fault occurs at *L*23 without any power source outage.

## IV. CONCLUSION

In this paper, an adaptive OC protection scheme is presented. This proposed scheme is useful for solar-based DC microgrids because of specific parameters of these grids, like the high rise time of fault current. The proposed method uses the communication level to access the shared information of the microgrid operation condition and adjacent OC relays' operation. Also, the proposed method uses clustering the minimum values of fault currents as setting groups to improve protection coordination and speed. Proven by real-time simulations, the proposed method is faster than the standard inverse scheme which is commonly used in OC protection.